\documentclass[useAMS,usenatbib]{mn2e}
\usepackage{epsfig}
\usepackage{deluxetable} 
\usepackage{url}
\usepackage{rotating}
\usepackage{amsmath}

\title[The awakening of the NLSy1 PKS 1502$+$036]{The awakening of the $\gamma$-ray narrow-Line Seyfert 1 galaxy PKS 1502$+$036}
\author[D'Ammando, Orienti, Finke, et al.]{F. D'Ammando$^{1,2}$\thanks{E-mail: dammando@ira.inaf.it}, M. Orienti$^{2}$, J. Finke$^{3}$, T. Hovatta$^{4}$, M. Giroletti$^{2}$, \newauthor W. Max-Moerbeck$^{5}$, T. J. Pearson$^{6}$, A. C. S. Readhead$^{6}$, R. A. Reeves$^{7}$, \noindent J. L. Richards$^{8}$\\
$^{1}$Dip. di Fisica e Astronomia, Universit\`a di Bologna, Via Ranzani 1, I-40127 Bologna, Italy \\
$^{2}$INAF - Istituto di Radioastronomia, Via Gobetti 101, I-40129 Bologna, Italy\\
$^{3}$U.S. Naval Research Laboratory, Code 7653, 4555 Overlook Ave. SW, Washington, DC 20375-5352, USA \\
$^{4}$Aalto University Mets\"ahovi Radio Observatory, Mets\"ahovintie 114, FI-02540 Kylm\"al\"a, Finland \\
$^{5}$Max-Planck-Institut f\"ur Radioastronomie, Auf dem H\"ugel 69, D-53121 Bonn, Germany  \\
$^{6}$Cahill Center for Astronomy and Astrophysics, California Institute of Technology 1200 E. California Blvd., Pasadena, CA 91125, USA \\
$^{7}$CePIA, Departamento de Astronomía, Universidad de Concepcion, Casilla 160-C, Concepcion, Chile \\
$^{8}$Department of Physics, Purdue University, 525 Northwestern Avenue, West Lafayette, IN 47907, USA}
\begin{document}

\date{Accepted. Received; in original form}

\maketitle

\label{firstpage}

\begin{abstract}
After a long low-activity period, a $\gamma$-ray flare from the narrow-line
Seyfert 1 PKS 1502$+$036 ($z=0.4089$) was detected by the Large Area Telescope (LAT) on board {\em Fermi} in 2015. On 2015
December 20 the source reached a daily peak flux, in the 0.1--300 GeV band, of
(93 $\pm$ 19)$\times$10$^{-8}$ ph cm$^{-2}$ s$^{-1}$, attaining a flux of (237
$\pm$ 71)$\times$10$^{-8}$ ph cm$^{-2}$ s$^{-1}$ on 3-hr time-scales, which
corresponds to an isotropic luminosity of (7.3 $\pm$ 2.1)$\times$10$^{47}$ erg
s$^{-1}$. The $\gamma$-ray flare was not accompanied by significant spectral
changes. We report on multi-wavelength radio-to-$\gamma$-ray observations of
PKS 1502$+$036 during 2008 August--2016 March by {\em Fermi}-LAT, {\em Swift},
{\em XMM-Newton}, Catalina Real-Time Transient Survey, and the Owens Valley Radio Observatory (OVRO). An increase in activity was observed on 2015 December 22 by {\em Swift} in
optical, UV, and X-rays. The OVRO 15 GHz light curve reached the highest flux
density observed from this source on 2016 January 12, indicating a delay of
about three weeks between the $\gamma$-ray and 15 GHz emission peaks. This
suggests that the $\gamma$-ray emitting region is located beyond the broad
line region. We compared the spectral energy distribution (SED) of an average
activity state with that of the flaring state. The two SED, with the
high-energy bump modelled as an external Compton component with seed photons
from a dust torus, could be fitted by changing the electron distribution
parameters as well as the magnetic field. The fit of the disc emission during
the average state constrains the black hole mass to values lower than 10$^8$
M$_{\odot}$. The SED, high-energy emission mechanisms, and $\gamma$-ray
properties of the source resemble those of a flat spectrum radio quasar. 
\end{abstract} 


\section{Introduction}

Relativistic jets are mainly produced by radio-loud active galactic nuclei
(AGN) such as blazars and radio galaxies hosted in giant elliptical galaxies \citep{blandford78}. The discovery by the Large Area Telescope (LAT) on-board
the {\em Fermi Gamma-Ray Space Telescope} of variable $\gamma$-ray
emission from narrow-line Seyfert 1 (NLSy1) galaxies revealed the presence of
a new class of AGN with relativistic jets \citep[e.g.,][]{abdo09a,abdo09b,dammando12,dammando15a}.
Considering that NLSy1 are usually hosted in spiral galaxies
\citep[e.g.,][]{deo06}, the presence of a relativistic jet in these sources
seems to be in contrast to the paradigm that the formation of relativistic
jets could happen in elliptical galaxies only \citep{boett02,marscher10}. This
finding poses intriguing questions about the nature of these objects and the
formation of relativistic jets. In particular, one of the debated properties
of NLSy1 is their relatively small black hole (BH) mass ($M_{BH}$ = 10$^{6-8}$
M$_{\odot}$) in comparison to blazars and radio galaxies. It was suggested
that the BH masses of NLSy1 are underestimated due either to the effect of
radiation pressure \citep{marconi08} or to projection effects
\citep{baldi16}. Higher BH masses than those derived by the virial method \citep[e.g.,][]{yuan08} are in agreement with the values estimated by
modelling the optical/UV data with a Shakura and Sunyaev disc spectrum \citep{calderone13}.
 
PKS\,1502$+$036 has been classified as a NLSy1 on the basis of its optical
spectrum: full width at half-maximum FWHM (H$\beta$) = (1082 $\pm$ 113) km
s$^{-1}$, [OIII]/H$\beta$ $\sim$ 1.1, and a strong Fe II bump \citep{yuan08}. Among the radio-loud NLSy1, PKS 1502$+$036 has one of the
highest radio-loudness values ($RL$ = 1549)\footnote{$RL$ being defined as the ratio between the 1.4\,GHz and 4400\,\AA\, rest-frame flux densities.}. The source exhibits a compact core-jet structure on pc-scales, with the radio emission
dominated by the core component, while the jet-like feature accounts for only
4 per cent of the total flux density \citep{orienti12,dammando13a}. Simultaneous
multi-frequency Very Large Array observations carried out at various epochs
showed substantial spectral and flux density variability. \citet{lister16}
analyzing the MOJAVE images of PKS 1502$+$036 collected during 2010--2013
found a jet component moving at sub-luminal speed (i.e., 1.1$\pm$0.4
$c$). Optical intra-day variability with a flux amplitude of about 10
per cent was reported for PKS 1502$+$036 by \citet{paliya13}. In infrared
bands, a variation of 0.1--0.2 mag in 180 days was observed by the {\em Wide-field Infrared Survey Explorer} \citep{jiang12}.

In the $\gamma$-ray energy band PKS 1502$+$036 was not detected in the 90's by
the Energetic Gamma-Ray Experiment Telescope (EGRET) on board the {\em Compton
  Gamma Ray Observatory} at E $>$ 100 MeV \citep{hartman99}. On the other
hand, the source has been included in the first, second, and third {\em
  Fermi}-LAT source catalogues \citep[1FGL, 2FGL, 3FGL;][]{abdo10,nolan12,acero15}. No significant increase of $\gamma$-ray flux was observed between
2008 August and 2012 November \citep{dammando13a}. In 2015 December, $\gamma$-ray flaring activity from PKS 1502$+$036 was detected on a daily time-scale by {\em Fermi}-LAT \citep{dammando15b}, confirmed at lower energies by {\em Swift} observations \citep{dammando15c}.

In this paper, we discuss the flaring activity of PKS\,1502$+$036 observed in 2015 December--2016 January in comparison to the 2008--2015
data collected from radio to $\gamma$ rays. The paper is organized as follows. In Section 2, we report the LAT data analysis and results. In Section 3 we present the results of the {\em Swift} and {\em XMM-Newton} observations. Optical data collected by the Catalina Real-Time Transient
Survey (CRTS) and radio data collected by the 40 m Owens Valley Radio
Observatory (OVRO) single-dish telescope are reported in Section 4. In Section
5, we discuss the properties and the modelling of the spectral energy
distribution (SED) of the source during an average activity state and the high
activity state. Finally, we draw our conclusions in Section 6. Throughout the
paper, a $\Lambda$ cold dark matter cosmology with $H_0$ = 71 km s$^{-1}$
Mpc$^{-1}$, $\Omega_{\Lambda} = 0.73$ and $\Omega_{\rm m} = 0.27$ \citep{komatsu11} is adopted. The corresponding luminosity distance at $z =0.4089$ \citep[i.e. the source redshift;][]{schneider10} is d$_L =  2220$\ Mpc. 
In the paper, the quoted uncertainties are given at the 1$\sigma$ level, unless otherwise stated, and the photon indices are
parametrized as $dN/dE \propto E^{-\Gamma}$ with $\Gamma$ = $s$+1 ($s$ is the spectral index).

\section{{\em Fermi}-LAT Data: analysis and results}
\label{FermiData}

The {\em Fermi}-LAT  is a pair-conversion telescope operating from 20 MeV to
$>$ 300 GeV. Further details about the {\em Fermi}-LAT are given in \citet{atwood09}. 

\begin{figure}
\centering
\includegraphics[width=7.5cm]{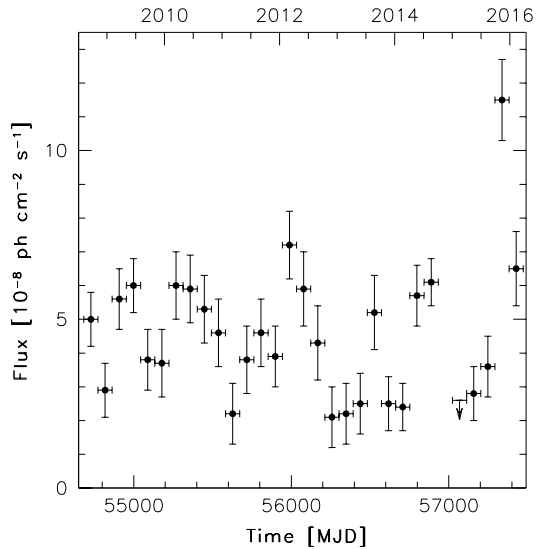}
\caption{Integrated flux light curve of PKS 1502$+$036 obtained in the
  0.1--300 GeV energy range during 2008 August 5--2016 March 24 (MJD
  54683--57471) with 90-day time bins. Arrow refers to 2$\sigma$ upper limit on the source flux. Upper limits are computed when TS $<$ 10.}
\label{Fig1}
\end{figure}

The LAT data used in this paper were collected from 2008 August 5  (MJD
54683) to 2016 March 24 (MJD 57471). During this time, the LAT instrument
operated almost entirely in survey mode. The Pass 8 data \citep{atwood13},
based on a complete and improved revision of the entire LAT event-level
analysis, were used. The analysis was performed with the
\texttt{ScienceTools} software package version v10r0p5. Only events belonging
to the `Source' class (\texttt{evclass=128}, \texttt{evtype=3}) were
used. We selected only events within a maximum zenith angle of 90 degrees to
reduce contamination from the Earth limb $\gamma$ rays, which are produced by cosmic rays interacting with the upper atmosphere. 
The spectral analysis was performed with the instrument response functions \texttt{P8R2\_SOURCE\_V6} using a binned maximum-likelihood method implemented
in the Science tool \texttt{gtlike}. Isotropic (`iso\_source\_v06.txt') and
Galactic diffuse emission (`gll\_iem\_v06.fit') components were used to model
the background \citep{acero16}\footnote{http://fermi.gsfc.nasa.gov/ssc/data/access/lat/\\BackgroundModels.html}. The normalization of both components was allowed to vary freely during the spectral fitting.

We analysed a region of interest of $30^{\circ}$ radius centred at the
location of PKS 1502$+$036. We evaluated the significance of the $\gamma$-ray
signal from the source by means of a maximum-likelihood test statistic (TS) defined as TS = 2$\times$(log$L_1$ - log$L_0$), where
$L$ is the likelihood of the data given the model with ($L_1$) or without ($L_0$) a point source at the position of PKS 1502$+$036
\citep[e.g.,][]{mattox96}. The source model used in \texttt{gtlike} includes all the point sources from the 3FGL catalogue that
fall within $40^{\circ}$ of PKS 1502$+$036. The spectra of these sources were parametrized by a power-law (PL), a log-parabola (LP), or a super
exponential cut-off, as in the 3FGL catalogue. We also included new candidates
within $10^{\circ}$ of PKS 1502$+$036 from a preliminary source list using 7 years of Pass 8 data.

A first maximum likelihood analysis was performed over the whole period to remove from the model the sources having TS $< 25$. A second maximum likelihood analysis was performed on the updated source model. In the fitting procedure, the normalization factors and the spectral parameters of the sources lying within 10$^{\circ}$ of PKS 1502$+$036 were left as free parameters. For the sources located between 10$^{\circ}$ and 40$^{\circ}$ from our target, we kept the normalization and the spectral shape parameters fixed to the values from the 3FGL catalogue.

Integrating over 2008 August 5--2016 March 24 the fit with a PL model, $dN/dE
\propto$ $(E/E_{0})^{-\Gamma_{\gamma}}$, as in the 3FGL catalogue, results in
TS = 1067 in the 0.1--300 GeV energy range, with an integrated average flux of (4.40 $\pm$ 0.21)$\times$10$^{-8}$ ph cm$^{-2}$ s$^{-1}$ and a photon index of $\Gamma_\gamma$ = 2.62 $\pm$ 0.04. The corresponding apparent isotropic $\gamma$-ray luminosity is (1.3$\pm$0.1)$\times$10$^{46}$ erg s$^{-1}$. 

Fig.~\ref{Fig1} shows the $\gamma$-ray light curve of PKS 1502$+$036 for 2008 August--2016 March using a PL model and 90-day time bins. For each time bin, the spectral parameters of PKS 1502$+$036 and all sources within 10$^{\circ}$ of it were frozen to the values resulting from
the likelihood analysis over the entire period. When TS $<$ 10, 2$\sigma$
upper limits were calculated. The statistical uncertainty in the fluxes are
larger than the systematic uncertainty \citep{ackermann12} and only the former
is considered in this paper.

Until September 2015, no significant variability was observed from PKS 1502$+$036 on a 90-day time-scale. The 0.1--300 GeV flux ranged between (2--7)$\times$10$^{-8}$ ph cm$^{-2}$ s$^{-1}$. An increase of activity was observed during 2015 September 29--December 25, when the source reached a 90-day averaged flux (0.1--300 GeV) of (11.5 $\pm$ 1.2)$\times$10$^{-8}$ ph
cm$^{-2}$ s$^{-1}$, a factor of 2.5 higher than the average $\gamma$-ray flux. Leaving the photon index of our target (and of all sources within
10$^{\circ}$ of our target) free to vary, the fit for PKS 1502$+$036 results in
TS = 155 and  a photon index $\Gamma_{\gamma}$ = 2.57$\pm$ 0.11, suggesting no
spectral variations during the high activity state. In order to test for curvature in the $\gamma$-ray spectrum of PKS 1502$+$036, an alternative spectral model to the PL, an LP, $dN/dE \propto$ $E/E_{0}^{-\alpha-\beta \, \log(E/E_0)}$, was used for the fit.
 We obtain a spectral slope $\alpha$ = 2.50 $\pm$ 0.14 at the reference energy $E_0$ = 246 MeV, a curvature parameter around the peak $\beta$ = 0.19 $\pm$ 0.13, and a TS = 156. We used a likelihood ratio test to check the PL model (null hypothesis) against the LP model (alternative hypothesis). These values may be compared by defining the curvature test statistic TS$_{\rm curve}$=TS$_{\rm LP}$--TS$_{\rm PL}$=1, meaning that we have no statistical evidence of a curved spectral shape.

\begin{figure}
\begin{center}
\includegraphics[width=7.5cm]{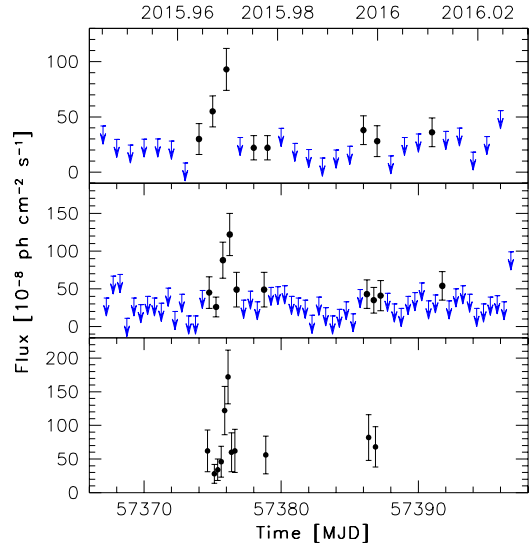}
\caption{\small{Integrated flux light curve of PKS 1502$+$036 obtained by {\em Fermi}-LAT in the 0.1--300 GeV energy range during 2015 December 11--2016 January 9 (MJD 57367--57396), with 1-day time bins (top panel), 12-h time bins (middle panel), and 6-h time bins (bottom panel). Arrows refer to 2$\sigma$ upper limits on the source flux. Upper limits are computed when TS $<$ 10. In the bottom panel upper limits are not shown.}}
\label{LAT_flare}
\end{center}
\end{figure}

In Fig.~\ref{LAT_flare} we show the light curve for the period 2015
December 11--2016 January 9 (MJD 57367--57396), with 1-day (top panel), 12-h
(middle panel), and 6-h (bottom panel) time bins. For each time bin, the
spectral parameters of PKS 1502$+$036 and all sources within 10$^{\circ}$ of
it were frozen to the values resulting from the likelihood analysis over the
entire period. In the following analysis of the sub-daily light curves, we fixed the flux of the diffuse emission components at the value obtained by fitting the data over the respective daily time-bins.

The daily peak of the emission was observed on 2015 December 20 (MJD 57376)
with a flux of (93 $\pm$ 19)$\times$10$^{-8}$ ph cm$^{-2}$ s$^{-1}$ in the
0.1--300 GeV energy range, 20 times higher than the average flux over the
whole period of {\em Fermi}-LAT observations. The corresponding apparent isotropic
$\gamma$-ray luminosity peak is (2.9 $\pm$ 0.6)$\times$10$^{47}$ erg
s$^{-1}$. Leaving the photon index free to vary the value obtained is $\Gamma$
= 2.54 $\pm$ 0.04, indicating that no significant spectral change is detected
during the high state on both daily and monthly time-scales. On a 12-h and a 6-h
time-scale the observed peak flux is (122 $\pm$ 28)$\times$10$^{-8}$ and (172
$\pm$ 40)$\times$10$^{-8}$ ph cm$^{-2}$ s$^{-1}$, respectively. The maximum
value on a 3-hr time-scale (light curve not shown) was observed on December 20 between 1:00 UT and 4:00 UT with a flux of (237 $\pm$ 71)$\times$10$^{-8}$ ph cm$^{-2}$ s$^{-1}$, corresponding to an apparent isotropic $\gamma$-ray luminosity of (7.3 $\pm$ 2.1)$\times$10$^{47}$ erg s$^{-1}$. 

After this main flux peak, a secondary peak was observed in the daily light curve on 2015 December 30, with a flux of (38 $\pm$ 13)$\times$10$^{-8}$ ph cm$^{-2}$ s$^{-1}$.
By means of the \texttt{gtsrcprob} tool, we estimated that the highest energy photon emitted by PKS 1502$+$036 (with probability $>$ 90\% of being
associated with the source) was observed on 2009 October 30 at a distance of 0\fdg05 from PKS 1502$+$036 with an energy of 21.1 GeV. Analyzing the
LAT data collected over 2008 August--2016 March in the 10-300 GeV energy range with a
PL the fit yielded a TS = 12, in agreement with the First {\em Fermi}-LAT
catalog of sources above 10 GeV \citep[1FHL;][]{ackermann13}, in which the source is
not reported. The 2$\sigma$ upper limit is 4.4$\times$10$^{-11}$ ph cm$^{-2}$ s$^{-1}$ (assuming a photon index
$\Gamma_{\gamma}$ = 3).

\begin{table*}
\caption{Log and fitting results of {\em Swift}-XRT observations of PKS 1502$+$036 using a PL model with $N_{\rm H}$ fixed to Galactic
  absorption.}
\begin{center}
\begin{tabular}{ccccc}
\hline
\multicolumn{1}{c}{Date (UT)} &
\multicolumn{1}{c}{MJD} &
\multicolumn{1}{c}{Net exposure time} &
\multicolumn{1}{c}{Photon index} &
\multicolumn{1}{c}{Flux 0.3--10 keV$^{\rm a}$} \\
\multicolumn{1}{c}{} &
\multicolumn{1}{c}{} &
\multicolumn{1}{c}{(sec)} &
\multicolumn{1}{c}{($\Gamma_{\rm\,X}$)} &
\multicolumn{1}{c}{($\times$10$^{-13}$ erg cm$^{-2}$ s$^{-1}$)} \\
\hline
2009-July-25 & 55037 & 4662 & $1.6 \pm 0.3$ & $5.1 \pm 0.7$ \\
2012-Apr-25  & 56042 & 4807 & $1.7 \pm 0.4$ & $4.0 \pm 0.7$ \\
2012-May-25  & 56072 & 4635 & $1.9 \pm 0.4$ & $4.0 \pm 0.7$ \\
2012-June-25 & 56103 & 5142 & $2.2 \pm 0.4$ & $3.8 \pm 0.8$ \\
2012-Aug-07/08 & 56146/7 & 4925 & $2.1 \pm 0.4$ & $4.7 \pm 0.8$ \\
2015-Dec-22 & 57378 & 2772 & $1.5 \pm 0.4$ & $6.7 \pm 0.7$ \\ 
2015-Dec-25 & 57381 & 2195 & $1.3 \pm 0.5$ & $6.4 \pm 0.6$ \\
2016-Jan-01 & 57388 & 1768 & $1.0 \pm 0.5$ & $11.6 \pm 0.6$ \\
2016-Jan-08 & 57395 & 2213 & $1.8 \pm 0.5$ & $6.3 \pm 0.7$ \\
2016-Jan-14 & 57401 & 2680 & $1.6 \pm 0.5$ & $7.9 \pm 0.6$ \\
\hline
\end{tabular}
\end{center}
$^{\rm a}$Unabsorbed flux
\label{XRT}
\end{table*}

\section{{\em Swift} and {\em XMM-Newton} observations}

\subsection{{\em Swift} data: analysis and results}
\label{SwiftData}

The {\em Swift} satellite \citep{gehrels04} carried out twelve observations of
PKS 1502$+$036 between 2009 July and 2016 January 22. The observations were
performed with all three instruments on board: the X-ray Telescope \citep[XRT;][0.2--10.0 keV]{burrows05}, the Ultraviolet/Optical Telescope \citep[UVOT;][170--600 nm]{roming05} and the Burst Alert Telescope \citep[BAT;][15--150 keV]{barthelmy05}.

The hard X-ray flux of this source turned out to be below the sensitivity of the BAT
instrument for such short exposures and therefore the data from this instrument will not be used.
Moreover, the source was not present in the {\em Swift} BAT 70-month hard X-ray catalogue \citep{baumgartner13}.

The XRT data were processed with standard procedures (\texttt{xrtpipeline
  v0.13.2}), filtering, and screening criteria by using the \texttt{HEAsoft}
package (v6.18). The data were collected in photon counting mode in all the
observations. The source count rate was low ($<$ 0.5 counts s$^{-1}$); thus
pile-up correction was not required. The data collected during 2012 August 7
and 8 were summed in order to have enough statistics to obtain a good spectral
fit. Source events were extracted from a circular region with a radius of 20
pixels (1 pixel $\sim$ 2.36 arcsec), while background events were extracted from a
circular region with radius of 50 pixels far away from the source
region. Ancillary response files were generated with \texttt{xrtmkarf}, and
account for different extraction regions, vignetting and point spread function
corrections. We used the spectral redistribution matrices v014 in the
Calibration data base maintained by \texttt{HEASARC}. Considering the low
number of photons collected ($<$ 200 counts) the spectra were rebinned with a
minimum of 1 count per bin and we used Cash statistics \citep{cash79}. 
We fitted the spectrum with an absorbed power-law using the photoelectric
absorption model \texttt{tbabs} \citep{wilms00}, with a neutral hydrogen
column density fixed to its Galactic value \citep[3.93$\times$10$^{20}$
  cm$^{-2}$;][]{kalberla05}. On 2016 January 22 the source was detected at a 2 $\sigma$ level with only 5 photons, therefore the spectrum is not fitted. The results of the fit are reported in Table~\ref{XRT}. The unabsorbed
fluxes in the 0.3--10 keV energy range are reported in Figs.~\ref{MWL} and ~\ref{MWL_zoom}.

During the {\em Swift} pointings, the UVOT instrument observed PKS 1502$+$036
in all its optical ($v$, $b$ and $u$) and UV ($w1$, $m2$ and $w2$) photometric
bands \citep{poole08,breeveld10}. We analysed the data using the
\texttt{uvotsource} task included in the \texttt{HEAsoft} package (v6.18). Source
counts were extracted from a circular region of 5 arcsec radius centred on
the source, while background counts were derived from a circular region of
10 arcsec radius in a nearby source-free region. The observed magnitudes are
reported in Table~\ref{UVOT}. Upper limits are calculated when the analysis
provided a detection  significance $<$ 3$\sigma$. The UVOT flux densities,
corrected for extinction using the E(B--V) value of 0.041 from
\citet{schlafly11} and the extinction laws from \citet{cardelli89}, are
reported in Figs.~\ref{MWL} and ~\ref{MWL_zoom}.

\begin{table*}
\caption{Results of the {\em Swift}-UVOT data for PKS 1502$+$036. Upper limits are calculated when the analysis provided a detection significance $<$ 3$\sigma$.}
\begin{center}
\begin{tabular}{cccccccc}
\hline
\multicolumn{1}{c}{Date (UT)} &
\multicolumn{1}{c}{MJD} &
\multicolumn{1}{c}{$v$} &
\multicolumn{1}{c}{$b$} &
\multicolumn{1}{c}{$u$} &
\multicolumn{1}{c}{$w1$} &
\multicolumn{1}{c}{$m2$} &
\multicolumn{1}{c}{$w2$} \\
\hline
2009-July-25 & 55037 & $>$ 18.63 & 19.16$\pm$0.25 & 18.38$\pm$0.18 & 18.40$\pm$0.16 & 18.22$\pm$0.16 & 18.44$\pm$0.12 \\
2012-Apr-25  & 56042 & 18.65$\pm$0.37 & 19.64$\pm$0.36 & 18.87$\pm$0.26 & 18.90$\pm$0.22 & 18.74$\pm$0.10 & 18.54$\pm$0.12 \\
2012-May-25  & 56072 & 18.68$\pm$0.34 & 19.46$\pm$0.28 & 18.64$\pm$0.06 & 18.29$\pm$0.14 & 18.45$\pm$0.16 & 18.25$\pm$0.09 \\
2012-June-25 & 56103 & $>$ 18.57 &  19.07$\pm$0.27 & 18.72$\pm$0.30 & 18.45$\pm$0.08 & 18.26$\pm$0.16 & 18.54$\pm$0.13 \\
2012-Aug-07  & 56146 & 18.33$\pm$0.27 & $>$19.65 & 19.28$\pm$0.35 & 18.51$\pm$0.16 & 18.74$\pm$0.19 & 18.39$\pm$0.10 \\
2012-Aug-08  & 56147 & 18.64$\pm$0.39 & 19.22$\pm$0.39 & 18.63$\pm$0.30 & 19.12$\pm$0.33 & 18.48$\pm$0.18 & 18.48$\pm$0.12 \\
2015-Dec-22  & 57378 & 18.31$\pm$0.24 & 18.45$\pm$0.16 & 18.09$\pm$0.16 & 17.83$\pm$0.10 & 17.88$\pm$0.07 & 18.14$\pm$0.08 \\ 
2015-Dec-25  & 57381 & $>$ 18.29 & $>$ 19.23 & 18.55$\pm$0.26 & 18.54$\pm$0.19 & 18.14$\pm$0.13 & 18.91$\pm$0.16 \\
2016-Jan-01  & 57388 & $>$ 17.99 & $>$ 19.17 & $>$ 20.12 & 18.38$\pm$0.19 & 18.39$\pm$0.17 & 18.41$\pm$0.12 \\
2016-Jan-08  & 57395 & $>$ 18.47 & 18.90$\pm$0.22 & 18.70$\pm$0.25 & 18.28$\pm$0.15 & 18.31$\pm$0.15 & 18.29$\pm$0.10 \\
2016-Jan-14  & 57401 & $>$ 18.16 & 19.10$\pm$0.30 & 18.27$\pm$0.21 & 18.17$\pm$0.16 & 18.35$\pm$0.17 & 18.17$\pm$0.08 \\
2016-Jan-22  & 57409 & $>$ 17.69 & $>$ 18.81 & 18.39$\pm$0.32 & 18.04$\pm$0.20 & 18.00$\pm$0.19 & 18.22$\pm$0.16 \\
\hline
\end{tabular}
\end{center}
\label{UVOT}
\end{table*}

\subsection{{\em XMM-Newton} data: analysis and results}
\label{XMMData}

{\em XMM-Newton} \citep{jansen01} observed PKS 1502$+$036 on 2012 August 7 for a total duration of 17 ks (observation ID 0690090101, PI: Foschini). The EPIC pn and the EPIC MOS cameras (MOS1 and MOS2) were operated in the full-frame mode. The data were reduced using the {\em XMM-Newton} Science Analysis System ({\small SAS v15.0.0}), applying standard event selection and filtering. Inspection of the background light curves showed that no strong flares were present during the observation, with good exposure times of 13.5, 16.8 and 16.9 ks for the pn, MOS1 and MOS2, respectively. For each of the detectors the source spectrum was extracted from a circular region of radius 32 arcsec centred on the source, and the background spectrum from a nearby region of radius 32 arcsec on the same chip. All the spectra were binned to contain at least 25 counts per bin to allow for $\chi^2$ spectral fitting. 

\begin{figure}
\begin{center}
\includegraphics[width=5.5cm, angle=-90]{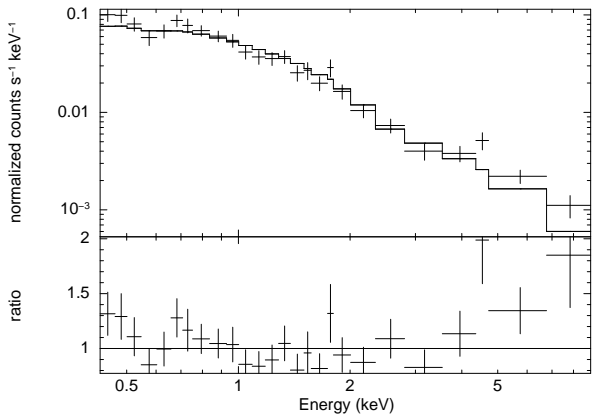}
\caption{Data and models (upper panel), and data--to--model ratio (lower panel) for the {\em XMM-Newton} EPIC-pn observation of PKS 1502$+$036 on
 2012 August 7 using a PL model.}
\label{XMM}
\end{center}
\end{figure}

All spectral fits were performed over the 0.4--10~keV energy range using {\small XSPEC v.12.9.0}. The energies of spectral features are quoted in
the source rest frame, while plots are in the observer frame. All errors are
given at the 90\% confidence level. Although we present only the fits to the
EPIC-pn, the results were cross-checked for consistency with the EPIC-MOS
spectra. Galactic absorption was included in all fits using the \texttt{tbabs}
model. The results of the fits are presented in Table~\ref{xmmfits}. A simple PL model is sufficient to describe the data, although some residuals
are present at low and high energies (Fig.~\ref{XMM}). The EPIC-pn flux
estimated in the 0.3--10 keV energy
range is (4.0$\pm$0.4)$\times10^{-13}$ erg\,cm$^{-2}$\,s$^{-1}$. There is no
significant detection of an Fe line in the spectrum, with a 90 per cent upper limit
on the equivalent width of 411 eV for a narrow emission line at 6.4~keV.

An improvement of the fit was obtained by using a broken power-law.
Applying an F-test the probability of obtaining such improvement by chance is
7$\times$10$^{-4}$. This can be an indication of the presence of both a
  soft X-ray excess below $\sim$2 keV and a relativistic jet component at
  higher energies, as observed in the $\gamma$-ray NLSy1 PMN J0948$+$0022 \citep{dammando14}.  However, the uncertainties on the photon index and flux
are quite large (Table~\ref{xmmfits}).

\section{Optical and radio observations}

\begin{table}
\begin{center}
\caption{Fits to the 0.4-10~keV {\em XMM-Newton} EPIC-pn spectra of PKS 1502$+$036. Galactic absorption was included in all fits.}
    \begin{tabular}{lll}
    \hline  Model  & Parameter  & Value \\
    \hline 
    Power law & \(\Gamma\) & \(1.9 \pm 0.1\) \\
          & Norm  & \(5.4\pm 0.3 \times10^{-5}\) \\ 
          & \(\chi^2\)/d.o.f.  & 46/39 \\ 
          \hline 
    Broken power law & \(\Gamma_1\) & \(2.2 \pm 0.3\) \\ 
          & \(E_{\rm{break}}~(\rm{keV})\) & \(1.6^{+1.1}_{-0.5}\) \\ 
          & \(\Gamma_2\) & \(1.5^{+0.2}_{-0.6}\)\\ 
          & Norm  & \(5.2^{+0.5}_{-0.6}\times10^{-5}\)  \\ 
          & \(\chi^2\)/d.o.f.  & 31/37 \\ 
          \hline    
    \end{tabular}\label{xmmfits}
    \end{center}     
\end{table}

\subsection{Optical CRTS data}\label{crts}

PKS 1502$+$036 has been monitored in 2008--2016 by the CRTS\footnote{http://crts.caltech.edu} \citep{drake09,djorgovski11},
using the 0.68 m Schmidt telescope at Catalina Station, AZ, and an unfiltered CCD. The typical cadence is to obtain four exposures separated by 10 min in a given night; this may be repeated up to four times per lunation, over a period of $\sim$6--7 months each year, while the field is
observable. Photometry is obtained using the standard Source-Extractor package \citep{bertin96}, and transformed from the unfiltered instrumental
magnitude\footnote{\url{http://nesssi.cacr.caltech.edu/DataRelease/FAQ2.html\#improve}} to Cousins $V$ by $V$ = $V_{\rm CSS}$ + 0.31($B-V$)$^{2}$ + 0.04. We averaged the values obtained during the same observing night. During the CRTS monitoring, the source showed a variability amplitude of 0.7 mag, changing between 18.89 and 18.19 mag. The CRTS flux densities, corrected for extinction using the E(B--V) value
of 0.041 from \citet{schlafly11} and the extinction laws from
\citet{cardelli89}, are reported in Fig.~\ref{MWL}.

\subsection{Radio OVRO data}\label{ovro}

\begin{figure*}
\centering
\includegraphics[width=11cm]{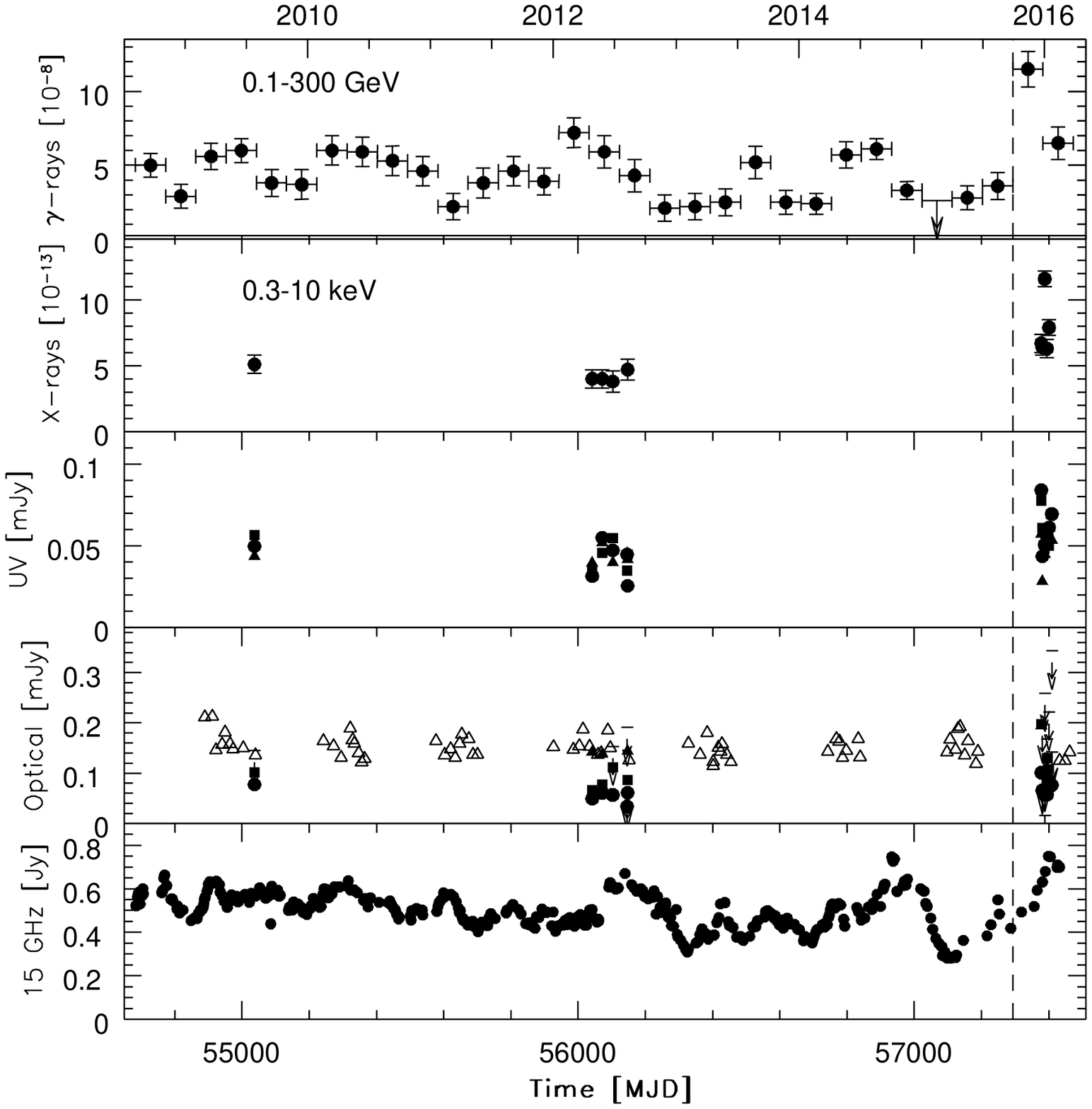}
\caption{Multi-frequency light curve for PKS 1502$+$036. The period covered is 2008 August--2016 March. The data sets were collected (from top to
bottom) by {\em Fermi}-LAT ($\gamma$ rays; in units of 10$^{-8}$ ph cm$^{-2}$
s$^{-1}$), {\em Swift}-XRT (0.3--10 keV; in units of 10$^{-13}$ erg cm$^{-2}$
s$^{-1}$), {\em Swift}-UVOT ($w1$, $m2$, $w2$ bands, shown as circles,
triangles, and squares, respectively;  in units of mJy), {\em Swift}-UVOT
($v$, $b$, $u$ bands, shown as filled triangles, circles, and squares,
respectively; in units of mJy), CRTS ($V$ band, shown as open triangles; in units of
mJy), and OVRO (15 GHz, in units of Jy). The vertical dashed line indicates the beginning of the period shown in detail in Fig.\ref{MWL_zoom}.}
\label{MWL}
\end{figure*}

\begin{figure*}
\centering
\includegraphics[width=11cm]{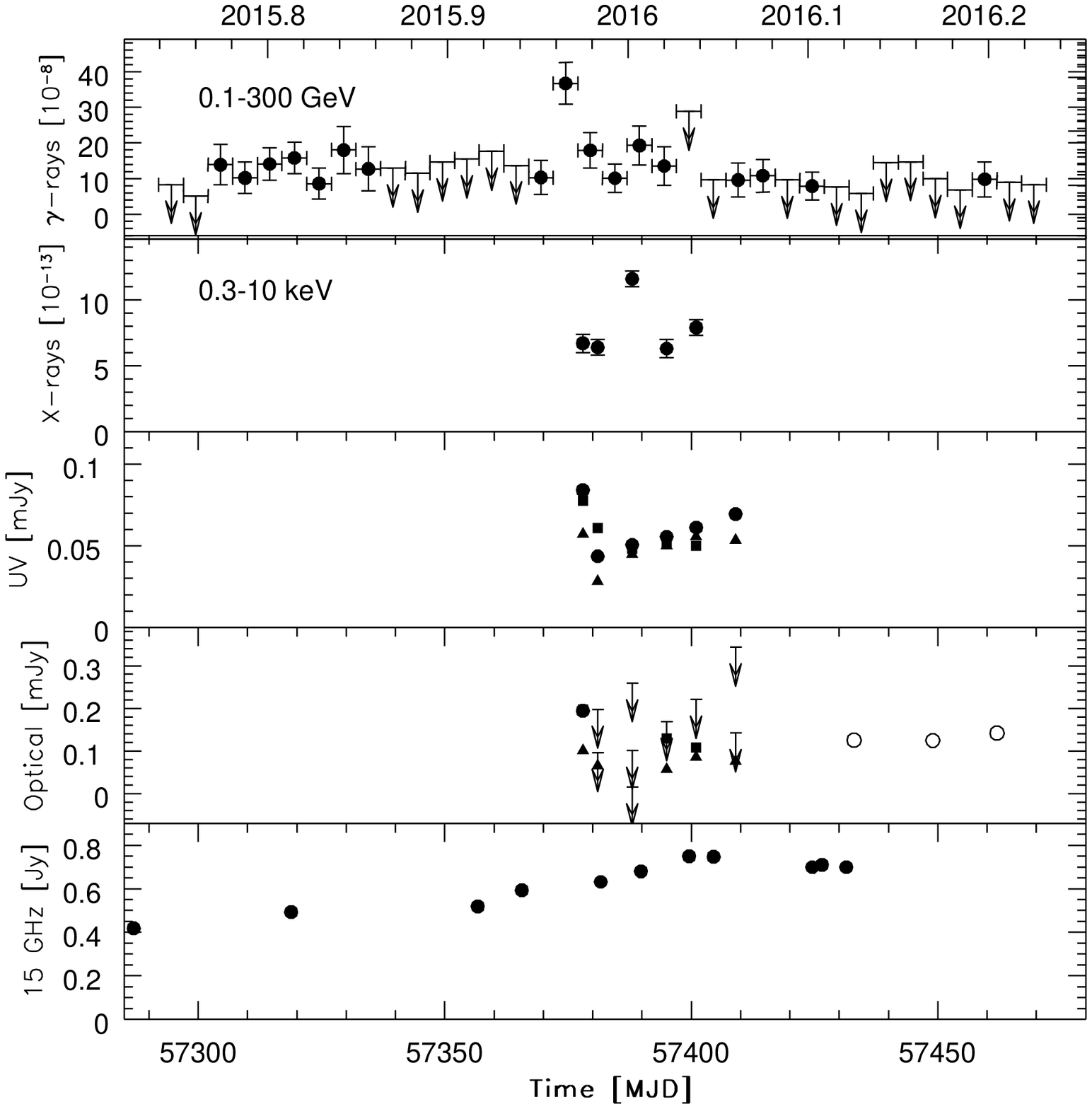}
\caption{Multi-frequency light curve for PKS 1502$+$036. The period covered is
  2015 September 29--2016 March 24. The data sets were collected (from top to
  bottom) by {\em Fermi}-LAT ($\gamma$ rays, with 5-day time bins; in units of
  10$^{-8}$ ph cm$^{-2}$ s$^{-1}$), {\em Swift}-XRT (0.3--10 keV; in units of
  10$^{-13}$ erg cm$^{-2}$ s$^{-1}$), {\em Swift}-UVOT ($w1$, $m2$, $w2$
  bands, shown as circles, triangles, and squares, respectively;  in units of
  mJy), {\em Swift}-UVOT ($v$, $b$, $u$ bands, shown as filled circles, triangles,
  and squares, respectively; in units of mJy), CRTS ($v$ band, shown as open
  circles; in units of mJy), and OVRO (15 GHz, in units of Jy).}
\label{MWL_zoom}
\end{figure*}

As part of an ongoing blazar monitoring programme, the OVRO 40 m radio
telescope has observed PKS 1502$+$036 at 15~GHz regularly during 2008--2016
\citep{richards11}. This monitoring programme includes over 1900 known and
likely $\gamma$-ray loud blazars above declination $-20^{\circ}$. The sources
in this programme are observed in total intensity twice per week with a 4~mJy
(minimum) and 3 per cent (typical) uncertainty in their flux densities. Observations
are performed with a dual-beam (each 2.5~arcmin FWHM) Dicke-switched system
using cold sky in the off-source beam as the reference. Additionally, the
source is switched between beams to reduce atmospheric variations. In 2014 May
a new pseudo-correlation receiver was installed on the 40 m telescope and the
fast gain variations are corrected using a 180 degree phase switch instead of
a Dicke switch. The performance of the new receiver is very similar to the old
one and no discontinuity is seen in the light curves. 
The
absolute flux density scale is calibrated using observations of 3C~286,
adopting the flux density (3.44~Jy) from \citet{baars77}. This results in
about a 5 per cent absolute scale uncertainty, which is not reflected in the plotted
errors. Flux densities at 15 GHz are reported in Figs.~\ref{MWL} and
\ref{MWL_zoom}. PKS 1502$+$036 was observed to be variable at 15 GHz during the OVRO
monitoring, with a flux density spanning from (282 $\pm$ 9) mJy (at MJD 57109) to (749 $\pm$ 10) mJy (at MJD 57399).

\section{Discussion}

\subsection{Multi-frequency variability}\label{Multifrequency}

Multi-wavelength follow-ups of $\gamma$-ray flares are crucial for
investigating possible connections between the $\gamma$-ray high activity
states and the variability observed in different energy bands. In Fig.~\ref{MWL}, we compare the $\gamma$-ray light curve obtained by {\em
  Fermi}-LAT during 2008 August--2016 March with the X-ray (0.3--10 keV), UV
($w1$, $m2$, and $w2$ filters), optical ($u$, $b$, and $v$ filters), and radio
(15 GHz) light curves obtained by {\em Swift}, CRTS and OVRO. A zoom of the
multi-frequency light curve during the high activity period (i.e. 2015 September
27--2016 March 24; MJD 57292--57471) is shown in Fig.~\ref{MWL_zoom}.

\noindent PKS 1502$+$036 showed no significant increase in the $\gamma$-ray activity during
2008 August--2015 September. The
average LAT spectrum accumulated over 2008 August--2016 March is well
described by a PL with a photon index of $\Gamma$ = 2.62 $\pm$ 0.04. This
photon index is similar to the average value observed for FSRQ  and steep
spectrum radio quasars during the first four years of {\em Fermi}-LAT operation
\citep[$\Gamma$ = 2.44 $\pm$ 0.20 and 2.42 $\pm$ 0.10,
  respectively;][]{ackermann15} rather than BL Lac objects ($\Gamma$ =
2.01 $\pm$ 0.25). The average apparent isotropic $\gamma$-ray luminosity of
PKS 1502$+$036 is (1.3 $\pm$ 0.1)$\times$10$^{46}$ erg s$^{-1}$ in the 0.1--300 GeV range, a typical value for a FSRQ \citep{ackermann15}. 

\noindent An increase in activity was observed in the period 2015 September 29--December 25, with a 90-day averaged flux (0.1--300 GeV) of (11.5 $\pm$ 1.2)$\times$10$^{-8}$ ph cm$^{-2}$ s$^{-1}$. No significant spectral change is detected during this high state, with a photon index of $\Gamma$ = 2.57 $\pm$ 0.11. The $\gamma$-ray flux increased by a factor of 4 in 5 days after mid-December 2015 (Fig.~\ref{MWL_zoom}).
On a daily time-scale the peak of activity was detected on 2015 December 20 (MJD 57376), with a flux of (93 $\pm$ 19)$\times$10$^{-8}$ ph cm$^{-2}$
s$^{-1}$, compatible with the daily peak values observed in the other $\gamma$-ray emitting NLSy1 \citep[e.g., SBS 0846$+$513, PMN J0948$+$0022, and 1H 0323$+$342;][]{dammando13b,dammando15d,carpenter13}. At the peak of the activity a maximum value of (7.3
$\pm$ 2.1)$\times$10$^{47}$ erg s$^{-1}$ was observed on a 3-h time-scale. Such a high value, together with the radio spectral variability
and the one-sided structure observed on parsec scale \citep{dammando13a}, suggests the presence of a relativistic jet with Doppler factors as large as in FSRQ.  

PKS 1502$+$036 was observed in X-rays by {\em Swift}/XRT in a bright state
during 2015 December--2016 January, with a flux a factor of 1.5--3 higher than
in the 2012 observations (Table~\ref{XRT}). X-ray spectra of NLSy1 are usually
characterized by a soft photon index, i.e. $\Gamma_{\rm X}$ $>$ 2
\citep[e.g.,][]{boller96}. The photon index during the high
activity period shows a moderate hardening and is always $\Gamma_{\rm X}$ $<$ 2, suggesting the presence of
an important contribution from the relativistic jet, as observed for other
$\gamma$-ray emitting NLSy1 \citep[e.g.,][]{dammando13b,dammando15d}. A very
hard photon index of $\Gamma_{\rm X}$ $\sim$1 was observed on 2016 January 1,
but no conclusive evidence can be drawn due to the large uncertainties. The
high flux observed during that observation is mainly due to the very hard
$\Gamma_{\rm X}$ estimated. 

\noindent The X-ray spectrum of PKS 1502$+$036 observed by {\em XMM-Newton} is
quite well reproduced by a single PL with photon index $\Gamma_{\rm X}$ = 1.9
$\pm$ 0.1, although some residuals at low and high energies are visible, favouring a broken power-law. 
The residuals at low energies may be the hint of soft X-ray excess, which is a usual feature in the X-ray spectrum of NLSy1
\citep{grupe10} and already detected in the {\em XMM-Newton} spectrum of the
$\gamma$-ray NLSy1 PMN J0948$+$0022 \citep{dammando13b}. In the same
way, the residuals observed at high energies may be an indication of the
presence of an Fe line in the X-ray spectrum. In this context the better fit obtained with a broken power-law model might indicate that the emission from the jet dominates above $\sim$2 keV, while a soft X-ray excess is present in the low-energy part of the X-ray spectrum.    
Unfortunately, during the {\em XMM-Newton} observation the flux was too low
and the observation time relatively short for detecting both the soft X-ray
excess and the Fe line. As for the $\gamma$-ray NLSy1 PKS 2004$-$447 \citep[see][]{orienti15}, deeper observations are needed for investigating in detail the presence of these features in the X-ray spectrum of PKS 1502$+$036.

During the {\em Swift} observation performed in 2015 December 22 the UV and optical emission reached a maximum soon after the $\gamma$-ray peak,
suggesting a common origin for the multi-frequency variability. The variability
amplitude (calculated as the ratio between the maximum and minimum flux
density) is 1.7, 3.0, 3.1, 3.4, 2.2, and 2.0 in the $v$, $b$, $u$, $w1$,
$m2$, and $w2$ bands, respectively. The decrease of amplitude variability in the $m2$ and $w2$
filters may be due to the presence of an accretion disc that dilutes the jet
emission in that part of the spectrum, as already observed for FSRQ
\citep[e.g.,][]{raiteri12} and the $\gamma$-ray NLSy1 PMN J0948$+$0022 \citep{dammando15d}.

The analysis of the 15 GHz light curve shows some flux density variability. In
particular, during 2008 August--2016 March PKS 1502$+$036 showed three outbursts: in 2012 April, 2014 October, and 2016 January. The first
two radio outbursts seem to be related to an increase of the $\gamma$-ray
flux by a factor of two at the beginning of 2012 and in mid-2014. After the 2014 October outburst
the flux density decreased until April 2015 when it reached the minimum
value. Then, it increased again for several months, and peaked on 2016 January 12,
 about three weeks after the peak of the $\gamma$-ray activity. The third outburst
showed the largest amplitude variability ($\sim$2.7) in the radio light curve,
reaching the highest flux density at 15 GHz for this source so far. 

\subsection{Radio and $\gamma$-ray connection in 2015--2016}

\noindent Following \citet{valtaoja99}, we estimate the variability time-scale
$\Delta t$ on the basis of the radio data. For $\Delta t$, we assume the time
interval between the minimum and maximum radio flux density of the outburst $|
\Delta S|$. This assumption implies that the minimum flux density corresponds
to a stationary underlying component and the variation is due to  a transient
component. Taking into consideration the time dilation due to the cosmological
redshift we find that the intrinsic time lag is $\Delta \tau = \Delta
t/(1+z)$, while the intrinsic flux density variation at the observed frequency
is $| \Delta S_{i}| =| \Delta S| \times (1+z)^{1-s}$ ($S_{\nu}$ $\propto$
$\nu^{-s}$). 

Following \citet{dammando13a}, we derive the rest-frame variability brightness temperature from

\begin{equation}
T'_{B} = \frac{2}{\pi k} \frac{|\Delta S| d_{L}^{2}}{\Delta t^{2} \nu^{2} {\rm (} 1+z {\rm )^{1+s}}}  \;\; .
\label{tbrest}
\end{equation}
 
\noindent where $k$ is the Boltzmann constant, $\nu$ is the observing frequency, and $s$ is the spectral index. During the outburst we have $\Delta S =$ 467 mJy and $\Delta t=$ 290 d. If in
equation~(\ref{tbrest}) we consider these values, and we assume $s=0.3$, i.e. the average value obtained by fitting the
optically thin spectrum \citep[see][]{dammando13a},  we obtain $T'_{B}$ $\sim$ 4.6$\times$10$^{12}$ K, which exceeds the value derived for the Compton catastrophe. Assuming that such a high value is due to Doppler boosting, we can estimate the variability Doppler factor $\delta_{\rm var}$, by means of:

\begin{equation}
\delta_{var} = \left( \frac{T'_{B}}{T_{int}} \right)^{1/(3+s)},
\label{dopplervar}
\end{equation}

\noindent where $T_{int}$ is the intrinsic brightness temperature. Assuming a
typical value $T_{int}$= 5$\times$10$^{10}$ K, as derived by
e.g.~\citet{readhead94,lahteenmaki99}, we obtain $\delta_{var}$ = 3.9. For the
radio outburst which occurred on 2012 July a $\delta_{var}$ = 6.6 was
obtained \citep{dammando13a}. If we consider as the radio outburst the period
MJD 57286--57399, we have $\Delta S =$ 331 mJy and $\Delta t=$ 113 d. In that
case we obtain $T'_{B}$ $\sim$ 2.1$\times$10$^{13}$ K, corresponding to a
$\delta_{var}$ = 6.2, comparable to the value obtained for the 2012 July
outburst. As a comparison, for the $\gamma$-ray emitting NLSy1 SBS 0846$+$513 and PMN J0948$+$0022 a variability Doppler factor of 11 and 8.7 was reported in \citet{dammando13b} and \citet{angelakis15}, respectively. 

We observed a delay of 22 days between the $\gamma$-ray and 15 GHz radio peak,
which corresponds to 15.6 days in the source's frame. 
By analyzing $\gamma$-ray and radio 15 GHz data for a sample of 183 sources
\citet{pushkarev10} found
that the $\gamma$-ray/radio delay ranges between 1 and 8 months in the observer's frame, with a peak at $\sim$1.2 months in the source's frame. 

\noindent Following \citet{pushkarev10}, we computed the de-projected distance between the $\gamma$-ray emitting region
($r_\gamma$) and the radius of the radio core ($r_\mathrm{c}$) at 15 GHz: 
\begin{equation}
\Delta r = r_\mathrm{c}-r_\gamma=\frac{\beta_\mathrm{app}c\Delta t_{\mathrm{R}-\gamma}^\mathrm{source}}{\sin\theta}\,,
\label{distance}
\end{equation}
where $\beta_\mathrm{app}$ is the apparent jet speed,
$\Delta$$t_{{\mathrm{R}-\gamma}}^\mathrm{source}$ is the radio to $\gamma$-ray
time delay in the source's frame ($\gamma$-ray leading), and $\theta$ is the
viewing angle. In the case of PKS 1502$+$036, by considering a
$\Delta$$t_{\mathrm{R}-\gamma}^\mathrm{source}$ = 15.6 days, and $\theta$ =
3$^{\circ}$, as assumed in \citet{abdo09a} and \citet{paliya16}, we obtain $\Delta$r = 7.6 $\times$10$^{17}$ $\times$ $\beta_\mathrm{app}$. Assuming $\beta_\mathrm{app}$ = 1.1, as estimated by \citet{lister16}, we obtain $\Delta$r = 8.4$\times$10$^{17}$ cm, i.e. 0.27 pc. However, this apparent velocity was derived during a period without significant $\gamma$-ray outbursts. If we assume a $\beta_\mathrm{app}$ = 10, similar to the values estimated for the $\gamma$-ray NLSy1 SBS 0846$+$513, 1H 0323$+$342, and PMN J0948$+$0022 \citep{lister16,dammando13b}, we obtain $\Delta$r = 7.6$\times$10$^{18}$ cm, i.e. 2.5 pc.

To evaluate the distance between the $\gamma$-ray emitting region and the jet base, we estimate the radius of the synchrotron self-absorbed radio 
core at 15 GHz. At any given frequency, the core is the surface where the optical depth $\tau$ is close to unity. Therefore, the apparent position of this unit-opacity surface depends on observing frequency \citep[e.g.,][]{konigl81}. In this scenario we can estimate the size of 15-GHz core, 
$\theta_{SSA}$ (mas) by:

\begin{equation}
\theta_{SSA} \sim f(s)^{5/4} \left(\frac{B}{G}\right)^{1/4} \left(\frac{\nu}{GHz}\right)^{-5/4} \left(\frac{S}{Jy}\right)^{1/2}  \left(\frac{1+z}{\delta}\right)^{1/4}
\label{ssa}
\end{equation}

\noindent where $B$ is the magnetic field, $S$ is the flux density at the
frequency $\nu$, $z$ is the redshift, $\delta$ is the Doppler factor and $f(s)$ is a function that depends slightly on $s$ \citep[e.g.,][]{kellermann81}. 

We assume that the magnetic field is in equipartition:

\begin{equation}
B \sim \left( \frac{c_{12}L}{V} \right)^{2/7}
\label{eq}
\end{equation}

\noindent where $L$ is the radio luminosity, $V$ the volume and $c12$ a constant that is tabulated in \citet{pacholcyz70} and depends on the spectral index and the upper and lower cut-off frequencies. By equating \ref{ssa} and \ref{eq} (see Appendix \ref{appendix}), we obtain the size that a source with the given flux density and frequency at the synchrotron self-absorption turnover must have to be in equipartition.

For PKS 1502$+$036 by considering the flux density reached at the peak of the
flare (0.749 Jy), and assuming $\delta = 3.9$, as derived by Eq. \ref{dopplervar}, and $s$ =
0.3, we obtain a projected 15-GHz core radius of $\sim$0.053 mas, i.e. $\sim$0.3 pc, which corresponds to a de-projected radius of $\sim$5.5
pc. Assuming  $\delta = 6.2$ we obtain a de-projected radius slightly smaller than the previous one.

It is worth noting that due to the uncertainties affecting the derivation of
some parameters, like the magnetic field and the Dopper factor, this value is
intended to provide an order of magnitude estimate of the radio core size, rather than an exact measurement. The result implies that the radio core at 15 GHz is located at parsec distance from the jet base, in agreement with other works \citep[e.g.,][]{pushkarev12,fuhrmann14,karamanavis16}.

By considering the distance between the $\gamma$-ray emitting region and the
15 GHz radio core estimated in Eq.~\ref{distance} for $\beta_{app}$ = 1.1 and
$\beta_{app}$ = 10, we locate the $\gamma$-ray emission region at a distance between 3.0 and 5.2 pc from the central BH. 
We can infer the broad line region (BLR) radius using the $R_{\rm\,BLR}$--$L$
relation from \citet{bentz13}. By using the luminosity at 5100 \AA\, L(5100) =
2.3$\times10^{44}$ erg s$^{-1}$ \citep{yuan08}, we obtain $R_{\rm\,BLR}$ = 1.4$\times$10$^{17}$ cm, i.e. $\sim$0.05 pc. 
Assuming that $\gamma$-ray and radio flares are related, and therefore separated by $\sim$3 weeks, this indicates that the $\gamma$-ray emitting region is also located at parsec scale distance from the jet base and beyond the BLR.

\subsection{SED modelling}\label{SED_model}

Estimates for the black hole mass for PKS 1502+036 span more than one order of magnitude. \citet{yuan08} reported $M_{BH}=4\times10^6 M_\odot$, based on a virial mass estimate of its optical spectrum. \citet{abdo09a}, \citet{calderone13}, and \citet{paliya16} found $M_{BH}=2\times10^7 M_\odot$,
$M_{BH}=2\times10^8 M_\odot$, and $M_{BH}=4.5\times10^7 M_\odot$, respectively, all from modeling the optical/UV SED as an accretion disc. Using
the optical spectrum from \citet{yuan08} and the BH mass scaling relation
from \citet{vestergaard06} we obtain $M_{BH}=1.4\times10^7 M_\odot$ for PKS
1502$+$036, in agreement with the value reported by \citet{abdo09a}, and
\citet{paliya16}. In the same way \citet{shaw12} using the Mg II line from the Sloan Digital Sky Server spectrum obtained a BH mass of 2.6$\times$10$^7$ $M_\odot$.  
It is worth noting that \citet{yuan08} used the BLR radius-luminosity relation from \citet{kaspi05} to estimate the mass of
the source. This is the reason why their estimate is inconsistent with our value. In our modeling here, we use $4.5\times10^7M_\odot$ following \citet{paliya16}.

\begin{figure} 
\centering
\vspace{2.0mm}
\includegraphics[width=7.0cm]{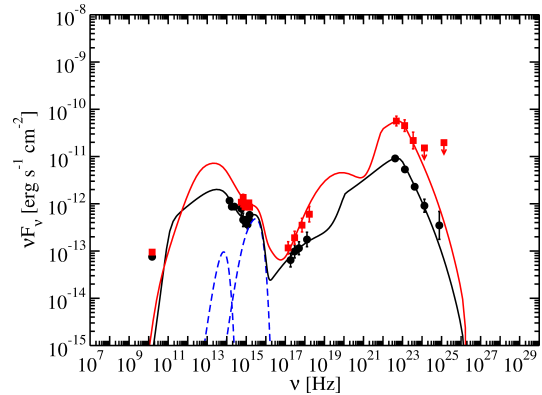}
\caption{Spectral energy distribution data (squares) and model fit (solid curve) of PKS\,1502$+$036 in flaring activity with the thermal emission components shown as dashed curves. The data points were collected by {\em Fermi}-LAT (2015 December 18--22), {\em Swift} (UVOT and XRT; 2015 December 22), and OVRO 40-m (2015 December 25). The SED in the average state reported is shown as circles and includes the {\em Fermi}-LAT spectrum from the 3FGL catalogue, the {\em Swift} (UVOT and XRT; 2012 April 25), and OVRO 40-m (2012 April 24) data.}
\label{SED}
\end{figure}

Two full, radio through $\gamma$-ray SED are shown in Figure \ref{SED}. The two SED represent an average state, and a high activity state. Aside from the thermal accretion disc emission, we fit the SED with a standard model for an emitting blob in a relativistic
jet aligned with our line of sight, producing synchrotron, synchrotron self-Compton (SSC), and external Compton (EC) emission. The synchrotron
component considered is self-absorbed below $10^{11}$\ Hz and thus cannot
reproduce the radio emission. This emission is likely from the superposition
of multiple self-absorbed jet components \citep{konigl81}.  We also included
thermal emission by an accretion disc and dust torus. See \citet{finke08} and
\citet{dermer09} for details on the model and formulae used. The variability
time was chosen to be about 1 day, consistent with the $\gamma$-ray light curve (Figure \ref{LAT_flare}). 

\begin{figure} 
\centering
\vspace{2.0mm}
\includegraphics[width=6.5cm]{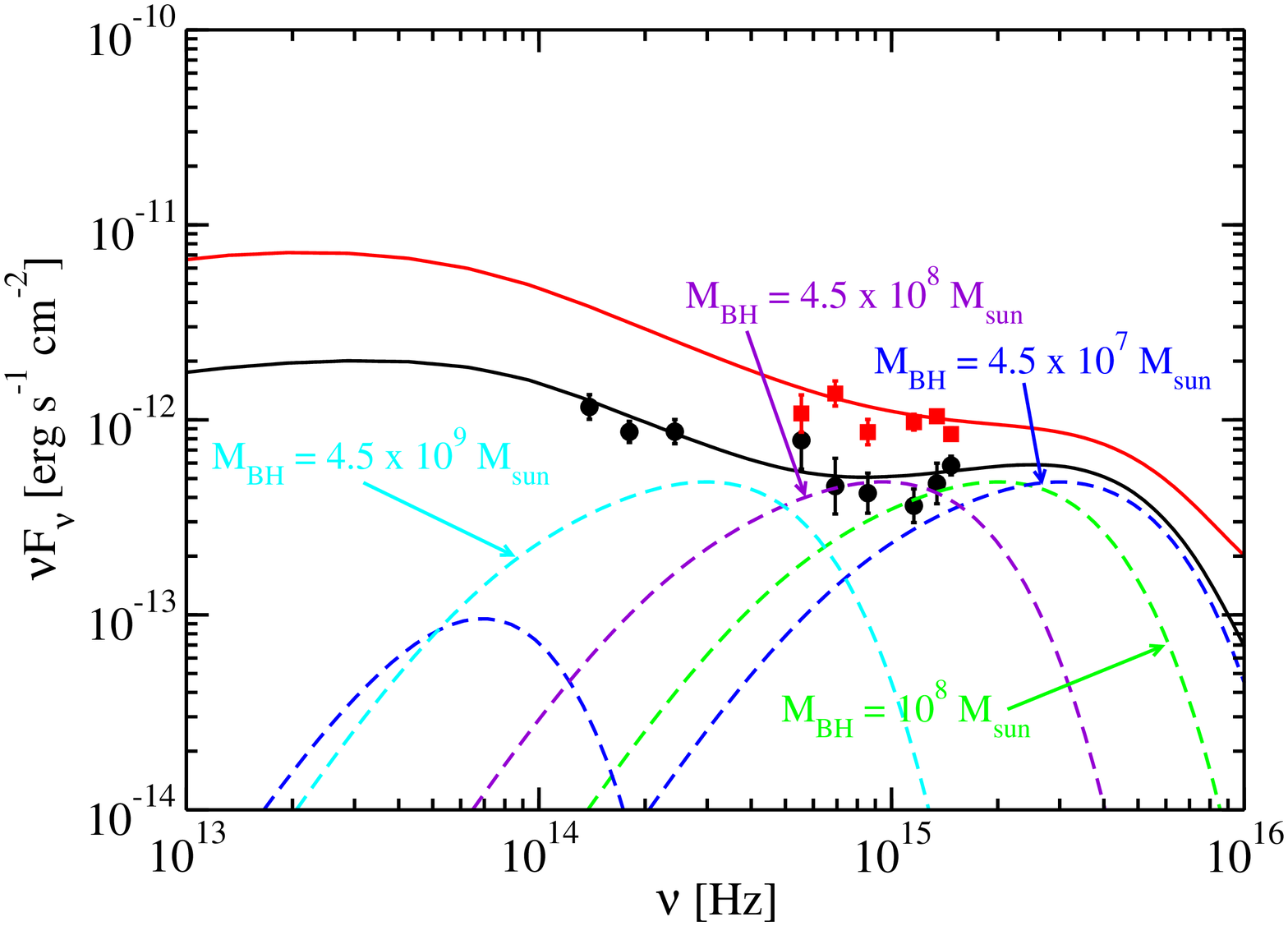}
\caption{Similar to Figure~\ref{SED}, but zoomed in on the optical/UV portion of the the average spectrum of PKS 1502$+$036. Model disc emission for different BH masses is shown as the dashed curves. The total (synchrotron + disc) emission is shown as the solid curves. Models with large BH mass do not provide an adequate fit to the UV data of 2012 April 25.}
\label{SED_zoom}
\end{figure}

We began by fitting the average state, starting with parameters used by
\citet{paliya16}. We only needed to modify their parameters slightly
to fit this SED; see the resulting model parameters in Table
\ref{table_fit}. The Doppler factor is similar to the value used in
\citet{abdo09a} for modelling an average SED of the source. 
In this model, most of the optical data are explained by synchrotron emission; the $\gamma$-ray data by EC emission; and the X-ray data by SSC emission. The EC model used is consistent with scattering of dust torus photons constrained by the sublimation radius \citep{nenkova08}.
We then attempted to fit the high state SED by varying only the electron distribution parameters from the average state.  In this, we
were not successful.  The $\gamma$-ray to optical ratio increases during the
flare. By considering that $F_{\gamma}$/$F_{\rm optical}$ $\propto$
($\delta/B)^2$, and assuming the optical emission is from synchrotron and the
$\gamma$-ray emission is from external Compton, either the Doppler factor or
the magnetic field have to change during the flare. This conclusion is fairly robust, since the optical, X-ray, and $\gamma$-ray data are contemporaneous, and the accretion disc contribution to the optical data is minimal.
 We chose to slightly modify the magnetic field in order to also fit the high
 state.  Unlike the modeling of \citet{paliya16}, we did not modify both the magnetic field and Doppler factor between the average and high state models, choosing instead to modify the minimum number of parameters between states. The optical/UV SED with several model curves for thermal emission from a Shakura-Sunyaev accretion disc is shown in Figure \ref{SED_zoom}. Within the assumed model, the mass is constrained to be $< 10^8 M_\odot$.

A Doppler factor $\delta$ $\sim$1--2, as inferred from the kinematic studies
of the MOJAVE images of the source \citep{lister16}, would not be able to fit
the SED. Such a low value would require a low magnetic field value to fit the
$\gamma$-ray and optical data. This would make the SSC very large, over-producing the X-ray emission observed, unless the emitting
region was made larger, which would conflict with the observed $\gamma$-ray
variability time-scale. This problem is similar to what is seen for TeV BL
Lacs. Bright TeV BL Lac objects have shown jet components with slow apparent speeds
\citep[e.g.,][]{piner10,lico12}, not compatible with Doppler factors inferred from
SED modeling \citep[e.g.,][]{abdo11}. It may be that the SED of PKS 1502$+$036 could be fit with a spine-layer model \citep{ghisellini05} or a decelerating jet model \citep{georganopoulos03}, both of which were proposed to resolve the TeV BL Lac Doppler factor discrepancy. However, we note that the SED of PKS 1502$+$036 does not resemble a TeV BL Lac, but an FSRQ, where there is no Doppler factor discrepancy.  We also note that no MOJAVE observations are collected during the high $\gamma$-ray activity period, missing a new superluminal jet component, if ejected.

The $\gamma$-ray luminosity and the photon index of PKS 1502$+$036 and indeed
its overall SED are similar to those of FSRQ, or low-synchrotron-peaked BL Lacs.  The two SED show a Compton dominance\footnote{Compton dominance is the ratio of the peak Compton
  luminosity to peak synchrotron luminosity.} $\sim$10. Note that in modelling
the SED of the FSRQ PKS 0537$-$441, several states could be reproduced by
varying only the electron distribution \citep{dammando13b}. In this way PKS
1502$+$036 is similar to the FSRQ PKS 2142$-$75 \citep{dutka13} and PKS
1424$-$418 \citep{buson14}, and the radio-loud NLSy1 SBS 0846$+$513
\citep{dammando13b} and PMN J0948$+$0022 \citep{dammando15d}. However, in PMN
J0948$+$0022 and SBS 0846$+$513, as well as for PKS 2142$-$75, the magnetic
field increased during the flare with respect to the low or average states. The behaviour of PKS 1502$+$036 seems to be opposite to the behaviour of these objects. Jet powers for both the average and high states of PKS 1502$+$036 are near equipartition between the electron and magnetic field energy density, with the electron energy density being slightly higher in both cases. 
 
\begin{table*}
\footnotesize
\begin{center}
\caption{Model parameters.}
\label{table_fit}
\begin{tabular}{lccc}
\hline
Parameter & Symbol & average state & high state \\
\hline
Redshift & 	$z$	& \multicolumn{2}{c}{0.409}  \\
Black hole Mass [M$_\odot]$ & $M_{BH}$ & \multicolumn{2}{c}{$4.5\times10^7$} \\
Bulk Lorentz Factor & $\Gamma$	& 17 & 17 \\
Doppler factor & $\delta_D$	& 17 & 17 \\
Magnetic Field [G]& $B$         & 0.35  & 0.27 \\
Variability Time-scale [s]& $t_v$       & \multicolumn{2}{c}{$7.5\times10^4$} \\
Comoving radius of blob [cm]& $R^{\prime}_b$ & 2.7$\times$10$^{16}$ & 2.7$\times$10$^{16}$ \\
\hline
Low-Energy Electron Spectral Index & $p_1$   & 2.3 & 2.3 \\
High-Energy Electron Spectral Index  & $p_2$ & 4.5 & 4.5 \\
Minimum Electron Lorentz Factor & $\gamma^{\prime}_{min}$  & $50$ & $500$ \\
Break Electron Lorentz Factor & $\gamma^{\prime}_{brk}$ & $1584$ & $1584$ \\
Maximum Electron Lorentz Factor & $\gamma^{\prime}_{max}$  & $1.5\times10^5$ & $1.5\times10^5$ \\
\hline
Disc luminosity [erg s$^{-1}$] & $L_{disk}$ & \multicolumn{2}{c}{$6.0\times10^{44}$} \\
Inner disc radius [$R_g$] & $R_{in}$ & \multicolumn{2}{c}{$6.0$} \\
Seed photon source energy density [erg cm$^{-3}$] & $u_{seed}$ & $7.2\times10^{-5}$ & $7.2\times10^{-5}$ \\
Seed photon source photon energy & $\epsilon_{seed}$ & $6.0\times10^{-7}$ & $6.0\times10^{-7}$ \\
Dust Torus luminosity [erg s$^{-1}$] & $L_{dust}$ & \multicolumn{2}{c}{$7.8\times10^{43}$} \\
Dust Torus radius [cm] & $R_{dust}$ & \multicolumn{2}{c}{$1.7\times10^{18}$} \\
Dust temperature [K] & $T_{dust}$ & \multicolumn{2}{c}{$1200$} \\
\hline
Jet Power in Magnetic Field [erg s$^{-1}$] & $P_{j,B}$ & $1.9\times10^{44}$ & $1.2\times10^{44}$ \\
Jet Power in Electrons [erg s$^{-1}$] & $P_{j,e}$ & $5.1\times10^{44}$ & $8.6\times10^{44}$ \\
\hline
\end{tabular}
\end{center}
\end{table*}

\section{Conclusions}

In this paper we reported on the observation by the {\em Fermi}-LAT of flaring
$\gamma$-ray activity from the NLSy1 PKS 1502$+$036 in 2015 December. On
2015 December 20 the source reached an apparent isotropic luminosity in the
0.1--300 GeV energy range of (2.9 $\pm$ 0.6)$\times$10$^{47}$ erg s$^{-1}$,
which is only a factor of 2--3 lower than those reached by the NLSy1 SBS 0846$+$513
and PMN J0948$+$0022 during a flare \citep{dammando13b,dammando15d}. On a 3-hr
time-scale the source reached a peak flux of (237 $\pm$ 71)$\times$10$^{-8}$
ph cm$^{-2}$ s$^{-1}$, corresponding to an apparent isotropic luminosity of
(7.3 $\pm$ 2.1)$\times$10$^{47}$ erg s$^{-1}$. 

The average photon index ($\Gamma_{\gamma}$ = 2.62 $\pm$ 0.04) and apparent
isotropic luminosity (L$_{\gamma}$ = 1.3 $\pm$ 0.1)$\times$10$^{46}$ erg
s$^{-1}$ estimated over 2008 August 5--2016 March 24 period are similar to the
values observed for FSRQ \citep[e.g.,][]{ackermann15}. No significant change
of the $\gamma$-ray spectrum was observed during the flare, with a photon index of $\Gamma_{\gamma}$ = 2.54 $\pm$ 0.04. 

In addition to the {\em Fermi}-LAT data, we presented multi-wavelength observations of PKS 1502$+$036 during the period 2008 August--2016 March including {\em Swift}, {\em XMM-Newton}, CRTS, and OVRO data. An increase of the activity was observed by {\em Swift} in X-rays,
UV, and optical on 2015 December 22, just a couple of days after the $\gamma$-ray peak, suggesting a common mechanism for the multi-frequency
variability during the flare. 

\noindent The source remained in a bright X-ray state during 2015 December--2016 January, with a photon index ranging between 1.0 and 1.8. These
values are harder than those observed in 2012, when the source had no significant high-energy outbursts. This suggests a dominant contribution of
the jet emission in the X-ray energy range during the high activity
period. The X-ray spectrum collected by {\em XMM-Newton} in 2012 was quite
well fit by a simple PL model, although some residuals are observed at low and
high energies.  These residuals hint at the presence of a soft X-ray excess and the Fe line, respectively. A better fit was obtained by
using a broken power-law model, suggesting the presence of two emission components in X-rays, but the uncertainties related to the spectral parameters are quite large. Deeper {\em XMM-Newton} observations are required for investigating these features in detail.

Flaring activity was also observed in the radio band. At 15 GHz the peak was detected on 2016 January 12, about three weeks after the $\gamma$-ray
peak. This radio flare may be the delayed counterpart of the $\gamma$-ray one due to opacity effects and the propagation of the shock along the jet. This suggests that the $\gamma$-ray emitting region is placed at $\sim$0.3 pc from the radio 15 GHz radius, and therefore at a distance between 3.0 and 5.2 pc from the central BH, well beyond the BLR.

We compared the broad-band SED of the 2016 flaring activity state with that
from an average state of PKS 1502$+$036 observed in 2012. Both the SED show a
Compton dominance  $\sim$10. This high value
indicates that the EC emission is the main mechanism for producing $\gamma$
rays, such as for FSRQ \citep[e.g.,][]{finke13},  confirming the
similarities between $\gamma$-ray emitting NLSy1 and FSRQ. The two SED,
with the high-energy bump modelled as an EC component of seed photons from a dust torus, could be
modelled by changing both the electron distribution parameters and the
magnetic field. An accretion disc is identified in the UV part of the spectrum
for the average activity state, with a luminosity of L$_{\rm\,disc}$ = 6$\times$10$^{44}$ erg s$^{-1}$. This value is lower than the luminosity
usually observed for FSRQ \citep[e.g.,][]{ghisellini14} as well as for the
$\gamma$-ray NLSy1 PMN J0948$+$0022 \citep{dammando15d}. On the other hand, no
evidence of thermal emission from the accretion disc has been observed for the
$\gamma$-ray NLSy1 SBS 0846$+$513 and PKS 2004$-$447, with a luminosity of the
accretion disc estimated to be very low \citep[10$^{42-43}$ erg s$^{-1}$;][]{dammando13b,orienti15}. 

No superluminal motion was observed in VLBI images during 2008--2012 \citep{dammando13a}, with only a sub-luminal component reported in
\citet{lister16}. This is in contrast to the radio spectral variability, the
one-sided structure, the observed $\gamma$-ray luminosity and the Doppler
factor estimated by SED modelling. This result resembles the `Doppler factor
crisis' observed in bright TeV BL Lacs. However, the SED of PKS 1502$+$036, in particular the high Compton dominance, does not resemble a TeV
BL Lac, but an FSRQ. Future VLBA monitoring of this NLSy1 during flaring activity periods may help to investigate this behaviour.

\noindent Assuming a BH mass of 4.5$\times$10$^7$ M$_{\odot}$, we obtain a
$L_{\rm\,disc}/L_{\rm\,Edd}$ = 0.1. Within the assumed model, the fit of the
disc emission of PKS 1502$+$036 during the average state constrains the black
hole mass to values lower than 10$^8$ M$_{\odot}$, and therefore to
$L_{\rm\,disc}/L_{\rm\,edd}$ $>$ 4$\times$10$^{-2}$, just above the threshold
between a radiatively efficient disc, as expected for FSRQ, and an inefficient one, as expected for BL Lacs \citep{ghisellini14}. The constraint of 10$^8$ M$_{\odot}$ obtained by modelling the optical/UV part of the spectrum of the source confirms that the radio-loud NLSy1, or at least the $\gamma$-ray emitting ones, are blazar-like sources with a BH mass between a few 10$^{7}$ M$_{\odot}$ and a few 10$^{8}$ M$_{\odot}$, therefore at the low end of the blazar distribution. The most powerful jets are found in luminous elliptical galaxies with very massive central BH, where the formation of the relativistic jets is usually triggered by strong merger activity \citep[e.g.,][]{sikora07,chiaberge15}. In this context it is unlikely that the $\gamma$-ray NLSy1 are hosted in disc/spiral galaxies like the other NLSy1 \citep[e.g.,][]{leontavares14}, but further observations of their host galaxies are needed to unravel the mystery.

\section*{Acknowledgements}

The {\em Fermi} LAT Collaboration acknowledges generous ongoing support from a number of agencies and institutes that have supported both the
development and the operation of the LAT as well as scientific data analysis. These include the National Aeronautics and Space Administration and the Department of Energy in the United States, the Commissariat \`a l'Energie Atomique and the Centre National de la Recherche Scientifique / Institut National de Physique Nucl\'eaire et de Physique des Particules in France, the Agenzia Spaziale Italiana and the Istituto Nazionale di Fisica Nucleare in Italy, the Ministry of Education, Culture, Sports, Science and Technology (MEXT), High Energy Accelerator Research Organization (KEK) and Japan Aerospace Exploration Agency (JAXA) in Japan, and the K.~A.~Wallenberg Foundation, the Swedish Research Council and the Swedish National Space Board in Sweden.
 
Additional support for science analysis during the operations phase is gratefully acknowledged from the Istituto Nazionale di Astrofisica in Italy and the Centre National d'\'Etudes Spatiales in France.

\noindent We thank the {\em Swift} team for making these observations possible, the
duty scientists, and science planners. The OVRO 40 m monitoring programme
is supported in part by NASA grants NNX08AW31G and NNX11A043G, and NSF grants AST-0808050 
and AST-1109911. The CSS survey is funded by the National Aeronautics and Space
Administration under Grant No. NNG05GF22G issued through the Science
Mission Directorate Near-Earth Objects Observations Programme.  The CRTS
survey is supported by the U.S.~National Science Foundation under
grants AST-0909182. Based on observations obtained with {\em XMM-Newton}, an
ESA science mission with instruments and contributions directly funded by ESA
Member States and NASA. Part of this work was done with the contribution of
the Italian Ministry of Foreign Affairs and Research for the collaboration
project between Italy and Japan. We thank S. Ciprini, J. Perkins, and the
anonymous referee for useful comments and suggestions.

\appendix

\section{The SSA radio core}\label{appendix}

In Section \ref{Multifrequency} we derived the size which a source with the given flux density and
frequency at the synchrotron self-absorption turnover must have to be in
equipartition. To do so we equate the magnetic field strength, $H_{\rm SSA}$, as
derived from the self-absorption expression with the equipartition magnetic field.
The former is computed by:

\begin{equation}
H_{\rm SSA} \sim f(s)^{-5} \left( \frac{\theta_{\rm SSA}}{{\rm mas}}\right)^{4}
\left( \frac{\nu}{{\rm GHz}}\right)^{5} \left( \frac{S}{{\rm Jy}}\right)^{-2}
\frac{\delta}{ (1+z)} \; {\rm G}
\label{ssa_2}
\end{equation}

\noindent where $f$(s) is a function that depends
slightly on $s$ \citep[e.g.][]{kellermann81}.
The equipartition magnetic field, $H_{\rm eq}$, is obtained by:

\begin{equation}
H_{\rm eq} \sim \left(c_{12} \frac{L}{V} \right)^{2/7}
\label{equi}
\end{equation}

\noindent where $L$ is the radio luminosity, $V$ the volume, and $c_{12}$ a value
tabulated in \citet{pacholcyz70}. The radio luminosity is obtained by

\begin{equation}
L = \frac{4 \pi D_{\rm L}^{2}}{(1+z)1-s} \int_{\nu_{1}}^{\nu_{2}} S(\nu) d\nu \,\, .
\label{luminosity}
\end{equation}

\noindent We approximate the volume of the source component
to a sphere which is homogeneously filled by
the relativistic plasma: 

\begin{equation}
V = \frac{4 \pi}{3} \left( \frac{D_{\rm L}}{(1+z)^2} \right)^3
\theta_{\rm eq}^3 \,\, .
\label{volum}
\end{equation}

\noindent If in Eq. \ref{equi} we consider Eqs. \ref{luminosity} and \ref{volum}, we
obtain:

\begin{equation}
\begin{split}
H_{\rm eq} & \sim 9 \times 10^{-7} c_{12}^{2/7} S_{0}^{2/7} \nu_{0}^{2s/7}
\delta^{\frac{-6 -2s}{7}} (1+z)^{\frac{10+2s}{7}} \\
 & \quad \times D_{\rm L}^{-2/7} \theta_{\rm eq}^{-6/7} c_{\nu}^{2/7}
\end{split}
\label{acca}
\end{equation}

\noindent where $S_{0}$ is the flux density in Jy at the frequency $\nu_{0}$ in Hz,
$D_{\rm L}$ is the luminosity distance in Mpc, $\theta_{\rm eq}$ is the radius in
mas, $H_{\rm eq}$ is the equipartition magnetic field in G, and $c_{\nu}$ is the
result of $\int_{\nu_{1}}^{\nu_{2}} \nu^{-s} d \nu$.\\  
If we equate Eqs. \ref{acca} and \ref{ssa_2}, and considering that $\theta_{\rm SSA} =
2\times \theta_{\rm eq}$, we have:

\begin{equation} 
\begin{split}
\theta_{\rm eq} & \sim 5 \times 10^{8} c_{12}^{1/17} c_{\nu}^{1/17} S_{0}^{8/17}
\nu_{0}^{\frac{2s - 35}{34}} \delta^{\frac{-13 -2s}{34}} \\
& \quad \times  (1+z)^{\frac{17+2s}{34}} D_{\rm L}^{1/17} \; .
\end{split}
\label{theta_fin}
\end{equation}


\begin{thebibliography}{99}

\bibitem[Abdo et al.(2009a)]{abdo09a} Abdo, A. A., et al. 2009a, ApJ, 707, L142

\bibitem[Abdo et al.(2009b)]{abdo09b} Abdo, A. A., et al. 2009b, ApJ, 699, 976

\bibitem[Abdo et al.(2010)]{abdo10} Abdo, A. A., et al. 2010, ApJS, 188, 405

\bibitem[Abdo et al.(2011)]{abdo11} Abdo, A. A., et al. 2011, ApJ, 736, 131

\bibitem[Acero et al.(2015)]{acero15}  Acero, F., et al. 2015, ApJS, 218, 23

\bibitem[Acero et al.(2016)]{acero16}  Acero, F., et al. 2016, ApJS, 223, 2

\bibitem[Ackermann et al.(2012)]{ackermann12} Ackermann, M., et al. 2012, ApJS, 203, 4

\bibitem[Ackermann et al.(2013)]{ackermann13} Ackermann, M., et al. 2013, ApJS, 209, 34

\bibitem[Ackermann et al.(2015)]{ackermann15} Ackermann, M., et al. 2015, ApJ, 810, 14

\bibitem[Angelakis et al.(2015)]{angelakis15} Angelakis, M., et al. 2015, A\&A, 575A, 55

\bibitem[Atwood et al.(2009)]{atwood09} Atwood, W. B., et al. 2009, ApJ, 697, 1071

\bibitem[Atwood et al.(2013)]{atwood13} Atwood, W. B., et al. 2013, 2012 Fermi Symposium proceedings - eConf C121028 (arXiv:1303.3514)

\bibitem[Baars et al.(1977)]{baars77} Baars, W. M., Genzel, R., Pauliny-Toth, I. I. K., Witzel, A. 1977, A$\&$A, 61, 99 

\bibitem[Baldi et al.(2016)]{baldi16} Baldi, R., Capetti, A., Robinson, A., Laor, A., Behar, E. 2016, MNRAS, 458, L69

\bibitem[Barthelmy et al.(2005)]{barthelmy05} Barthelmy, S. D., et al. 2005, Space Sci. Rev., 120, 143 

\bibitem[Baumgartner et al.(2013)]{baumgartner13} Baumgartner, W. H., Tueller, J., Markwardt, C. B., Skinner, G. K., Barthelmy, S., Mushotzky, R. F., Evans, P., Gehrels, N. 2013, ApJS, 207, 19
 
\bibitem[Bentz et al.(2013)]{bentz13} Bentz, M. C., et al. 2016, ApJ, 767, 149
 
\bibitem[Bertin $\&$ Arnouts(1996)]{bertin96} Bertin, E., $\&$ Arnouts, S. 1996, A\&AS,  117, 393

\bibitem[Blandford $\&$ Rees(1978)]{blandford78} Blandford, R.~D., \& Rees, M.~J. 1978, in Pittsburgh Conference on BL Lac Objects, ed A.~M. Wolfe,
  University Pittsburgh Press, 328 

\bibitem[Boller et al.(1996)]{boller96} Boller, T., Brandt, W. N., Fink, H. 1996 A\&A, 305, 53

\bibitem[B{\"o}ttcher \& Dermer(2002)]{boett02} B{\"o}ttcher, M., \& Dermer, C.~D.\ 2002, ApJ, 564, 86

\bibitem[Breeveld et al.(2010)]{breeveld10} Breeveld, A. A., et al. 2010, MNRAS, 406, 1687

\bibitem[Burrows et al.(2005)]{burrows05} Burrows, D. N., et al. 2005, Space
  Sci. Rev., 120, 165  

\bibitem[Buson et al.(2014)]{buson14} Buson, S., et al. 2014, A\&A, 569, A40

\bibitem[Calderone et al.(2013)]{calderone13} Calderone, G., Ghisellini, G., Colpi, M., Dotti, M. 2013, MNRAS, 431, 210 

\bibitem[Cardelli et al.(1989)]{cardelli89}  Cardelli, J. A., Clayton, G. C., Mathis, J. S. 1989, ApJ, 345, 245

\bibitem[Carpenter et al.(2013)]{carpenter13} Carpenter, B., Ojha, R. 2013, the Astronomer's Telegram, 5344

\bibitem[Cash(1979)]{cash79} Cash, W. 1979, ApJ, 228, 939
 
\bibitem[Chiaberge et al.(2015)]{chiaberge15} Chiaberge, M., Gilli, R., Lotz, J. M., Norman, C. 2015, ApJ, 806, 147

\bibitem[D'Ammando et al.(2012)]{dammando12} D'Ammando, F., et al . 2012, MNRAS, 426, 317

\bibitem[D'Ammando et al.(2013a)]{dammando13a} D'Ammando, F., et al. 2013a, MNRAS, 433, 952

\bibitem[D'Ammando et al.(2013b)]{dammando13b} D'Ammando, F., et al. 2013b, MNRAS, 436, 191

\bibitem[D'Ammando et al.(2013c)]{dammando13_0537} D'Ammando, F., et al. 2013c, MNRAS, 431, 2481

\bibitem[D'Ammando et al.(2014)]{dammando14} D'Ammando et al. 2014, MNRAS, 438, 3521

\bibitem[D'Ammando et al.(2015a)]{dammando15a} D'Ammando, F., Orienti, M., Larsson, J. , Giroletti, M.,  2015a, MNRAS, 452, 520

\bibitem[D'Ammando et al.(2015b)]{dammando15b} D'Ammando, F., Ciprini, S. 2015b, the Astronomer's Telegram, 8447, 1

\bibitem[D'Ammando (2015c)]{dammando15c} D'Ammando, F. 2015c, the Astronomer's Telegram, 8450, 1

\bibitem[D'Ammando et al.(2015d)]{dammando15d} D'Ammando, F., et al. 2015d, MNRAS, 446, 2456

\bibitem[Deo et al.(2006)]{deo06} Deo, R. P., Crenshow, D. M., Kraemer, S. B. 2006, AJ, 132, 321

\bibitem[Dermer et al.(2009)]{dermer09} Dermer, C.~D., Finke, J.~D., Krug, H., B\"ottcher, M.\ 2009, ApJ, 692, 32

\bibitem[Djorgovski et al.(2011)]{djorgovski11} Djorgovski, S.G., et al. 2011, in The First Year of MAXI: Monitoring Variable X-ray Sources, eds. T. Mihara \& N. Kawai. JAXA Special Publication, Tokio (arXiv:1102.5004)

\bibitem[Drake et al.(2009)]{drake09} Drake, A. J., et al. 2009, ApJ, 696, 870

\bibitem[Dutka et al.(2013)]{dutka13} Dutka, M.~S., et al.\ 2013, ApJ, 779, 174

\bibitem[Finke et al.(2008)]{finke08} Finke, J.~D., Dermer, C.~D., B\"ottcher, M.\ 2008, ApJ, 686, 181

\bibitem[Finke(2013)]{finke13} Finke, J.~D. 2013, ApJ, 763, 134

\bibitem[Fuhrmann et al.(2014)]{fuhrmann14} Fuhrmann, L., et al. 2014, MNRAS, 441, 1899

\bibitem[Gehrels et al.(2004)]{gehrels04} Gehrels, N., et al. 2004, ApJ, 611, 1005 

\bibitem[Georganopoulos \& Kazanas(2003)]{georganopoulos03} Georganopoulos, M., Kazanas, D. 2003, ApJ, 594, L27

\bibitem[Ghisellini et al.(2005)]{ghisellini05} Ghisellini, G., Tavecchio, F., Chiaberge, M. 2005, A\&A, 432, 401

\bibitem[Ghisellini et al.(2014)]{ghisellini14} Ghisellini, G., Tavecchio, F., Maraschi, L., Celotti, A., Sbarrato, T. 2014, Nature, 515, 376

\bibitem[Grupe et al.(2010)]{grupe10} Grupe, D., Komossa, S. Leighly, K. M., Page, K. L. 2010, ApJS, 187, 64

\bibitem[Hartman et al.(1999)]{hartman99} Hartman, R. C., et al. 1999, ApJS 123, 79

\bibitem[Kalberla et al.(2005)]{kalberla05} Kalberla, P. M. W., Burton, W. B., Hartmann, D., Arnal, E. M., Bajaja, E., Morras, R., P{\"o}ppel, W. G. L 2005, A$\&$A, 440, 775

\bibitem[Karamanavis et al.(2016)]{karamanavis16} Karamanavis, V., et al. 2016, A\&A, 590A, 48

\bibitem[Kaspi et al.(2005)]{kaspi05} Kaspi, S., Maoz, D., Netzer, H., Peterson, B. M., Vestergaard, M., Jannuzi, B. T. 2005, ApJ, 629, 61

\bibitem[Kellermann et al.(1981)]{kellermann81} Kellermann, K. I., \& Pauliny-Toth, I. I. K. 1981, ARA\&A, 19, 373

\bibitem[Komatsu et al.(2011)]{komatsu11} Komatsu, E., et al. 2011, ApJS, 192, 18

\bibitem[K{\"o}nigl(1981)]{konigl81} Konigl, A.\ 1981, ApJ, 243, 700

\bibitem[Jansen et al.(2001)]{jansen01}  Jansen F. et al., 2001, A\&A, 365, L1

\bibitem[Jiang et al.(2012)]{jiang12} Jiang, N., et al. 2012, ApJL, 759, L31

\bibitem[L\"ahteenm\"aki \& Valtaoja(1999)]{lahteenmaki99} L\"ahteenm\"aki, A., \& Valtaoja, E. 1999, ApJ, 521, 493

\bibitem[Leon Tavares et al.(2014)]{leontavares14} Leon Tavares, J., et al. 2014, ApJ, 795, 58

\bibitem[Lico et al.(2012)]{lico12} Lico, R., et al. 2012, A\&A, 545, 117

\bibitem[Lister et al.(2016)]{lister16} Lister, M. L., et al. 2016, AJ, in press

\bibitem[Marconi et al.(2008)]{marconi08} Marconi, A., et al. 2008, ApJ, 678, 693

\bibitem[Marscher(2010)]{marscher10} Marscher, A. 2010, in Lecture Notes in Physics 794, ed. T. Belloni (Berlin:Springer), 173

\bibitem[Mattox et al.(1996)]{mattox96} Mattox, J. R., et al. 1996, ApJ, 461, 396 

\bibitem[Nenkova et al.(2008)]{nenkova08} Nenkova, M., Sirocky, M.~M., Nikutta, R., Ivezi{\'c}, {\v Z}., \& Elitzur, M.\ 2008, ApJ, 685,
160

\bibitem[Nolan et al.(2012)]{nolan12} Nolan, P., et al. 2012, ApJS, 199, 31 

\bibitem[Orienti et al.(2012)]{orienti12} Orienti, M., D’Ammando, F., Giroletti, M., on behalf of the Fermi LAT Collaboration 2012, in Fermi and Jansky: Our evolving understanding of AGN, eConf C1111101 (arXiv:1205.0402) 

\bibitem[Orienti et al.(2015)]{orienti15} Orienti, M., D'Ammando, F., Larsson, J., Finke, J., Giroletti, M., Dallacasa, D., Isacsson, T., Stoby Hoglund, J. 2015, MNRAS, 453, 4037

\bibitem[Pacholcyzk(1970)]{pacholcyz70} Pacholczyk, A.G. 1970, Radio astrophysics. Nonthermal processes in galactic and extragalactic sources (San Francisco: W. H. Freeman \& Co Ltd)

\bibitem[Paliya et al.(2013)]{paliya13} Paliya, V. S., Stalin, C. S., Kumar, B., Kumar, B., Bhatt, V. K., Pandey, S. B., Yadav, R. K. S. 2013, MNRAS, 428, 2450

\bibitem[Paliya \& Stalin(2016)]{paliya16} Paliya, V.~S., \& Stalin, C.~S.\ 2016, ApJ, 820, 52

\bibitem[Piner et al.(2010)]{piner10} Piner, B. G., Pant, N., Edwards, P. G. 2010, ApJ, 723, 1150

\bibitem[Poole et al.(2008)]{poole08} Poole, T. S., et al. 2008, MNRAS, 383, 627

\bibitem[Pushkarev et al.(2010)]{pushkarev10} Pushkarev, A. B., Kovalev, Y. Y., Lister, M. L. 2010, ApJ, 722, L7

\bibitem[Pushkarev et al.(2012)]{pushkarev12} Pushkarev, A. B., Hovatta, T., Kovalev, Y. Y., Lister, M. L., Lobanov, A. P., Savolainen, T., Zensus, J. A, A\&A, 545, 113

\bibitem[Raiteri et al.(2012)]{raiteri12} Raiteri, C. M., et al. 2012, A\&A, 545, A48

\bibitem[Readhead(1994)]{readhead94} Readhead, A. C. S. 1994, ApJ, 426, 51

\bibitem[Richards et al.(2011)]{richards11} Richards, J. L., et al. 2011, ApJS, 194, 29 

\bibitem[Roming et al.(2005)]{roming05} Roming, P. W. A., et al. 2005, Space Sci. Rev., 120, 95 

\bibitem[Schlafly \& Finkbeiner(2011)]{schlafly11} Schlafly, E. F. \& Finkbeiner, D. P. 2011, ApJ, 737, 103

\bibitem[Schneider et al.(2010)]{schneider10} Schneider, D. P., et al. 2010, AJ, 139, 2360
                                   
\bibitem[Shaw et al.(2012)]{shaw12} Shaw, M. et l. 2012, ApJ, 748, 49

\bibitem[Sikora et al.(2007)]{sikora07}  Sikora, M., Stawarz, L., Lasota, J.-P. 2007, ApJ, 658, 815

\bibitem[Valtaoja et al.(1999)]{valtaoja99} Valtaoja, E., L\"ahteenm\"aki, A., Ter\"asranta, H., Lainela, M. 1999, ApJS, 120, 95

\bibitem[Vestergaard \& Peterson(2006)]{vestergaard06} Vestergaard, M., \& Peterson, B. M. 2006, ApJ, 641, 689

\bibitem[Wilms et al.(2000)]{wilms00} Wilms, J., Allen, A., McCray, R. 2000, ApJ, 542, 914 

\bibitem[Yuan et al.(2008)]{yuan08} Yuan, W., et al. 2008, ApJ, 685, 801

\end{thebibliography}
\end{document}